\documentclass[letters,fleqn,usenatbib]{mnras}

\usepackage{newtxtext,newtxmath}
\usepackage[T1]{fontenc}

\DeclareRobustCommand{\VAN}[3]{#2}
\let\VANthebibliography\thebibliography
\def\thebibliography{\DeclareRobustCommand{\VAN}[3]{##3}\VANthebibliography}

\usepackage{graphicx}	% Including figure files
\usepackage{amsmath}	% Advanced maths commands
\usepackage{hyperref}
\usepackage{orcidlink}
\usepackage{xspace}
\usepackage{textcase}
\usepackage{xstring}
\newcommand{\citepaperI}{\protect\hyperlink{cite.Klein2024_pendulum}{Paper I}\xspace}
\newcommand{\citepaperII}{\protect\hyperlink{cite.Klein2024_CDA}{Paper II}\xspace}

\title[KLCs with a precessing Quadrupole is a simple pendulum]{Hierarchical Three-Body Problem at High Eccentricities = Simple Pendulum\\III: Precessing Quadrupole}

\author[Y. Y. Klein and B. Katz]{
Ygal Y. Klein\orcidlink{0009-0004-1914-5821}\thanks{E-mail: ygalklein@gmail.com (YK)}
and Boaz Katz\orcidlink{0000-0003-0584-2920}\thanks{E-mail: boaz.katz@weizmann.ac.il (BK)}
\\
Dept. of Particle Phys. \& Astrophys., Weizmann Institute of Science,
 Rehovot 76100, Israel
}

\date{Accepted XXX. Received YYY; in original form ZZZ}

\pubyear{\the\year{}}

\begin{document}
\label{firstpage}
\pagerange{\pageref{firstpage}--\pageref{lastpage}}
\maketitle

\begin{abstract}
  The very long-term evolution of the hierarchical restricted three-body problem with a slightly aligned precessing quadrupole potential is investigated analytically and solved for both rotating and librating Kozai-Lidov cycles (KLCs) with high eccentricities.  We describe the finding of a striking similarity between librating and rotating KLCs for some range of precession rates. We show that the main effect occurs in both categories when the KLC frequency is equal to the precession rate of the perturbing potential. We solve the resonant dynamics analytically and show that it is equivalent to a simple pendulum model allowing us to map the strikingly rich structures that arise for precession rates similar to the Kozai-Lidov timescale (ratio of a few) and explain the similarity and when it vanishes. Additionally, we show that the regular KLCs at high eccentricities can also be described using a simple pendulum.
\end{abstract}

\begin{keywords}
gravitation-celestial mechanics-planets and satellites: dynamical evolution and stability-stars: multiple: close
\end{keywords}

%%%%%%%%%%%%%%%%%%%%%%%%%%%%%%%%%%%%%%%%%%%%%%%%%%

%%%%%%%%%%%%%%%%% BODY OF PAPER %%%%%%%%%%%%%%%%%%

\section{Introduction}
In this Letter we extend and simplify the analytical study and solution for the long-term evolution of a test particle in a Keplerian orbit perturbed by a slightly aligned precessing quadrupole potential presented for librating Kozai-Lidov Cycles (KLCs) in \citet{klein2023,klein2024}.
The system under consideration involves a test particle orbiting a central mass $M$ on a Keplerian orbit with semimajor axis $a$ that is perturbed by a precessing external quadrupole potential in the (time dependent) direction
\begin{equation}
    \mathbf{\hat{j}}_{\text{outer}}=\left(\sin \alpha \cos\left(\beta\tau\right), \ -\sin \alpha \sin\left(\beta\tau\right), \ \cos \alpha\right)
    \label{eq:jOuter_as_a_function_of_tau}
\end{equation}
which has a constant inclination $\alpha$ with respect to the \textit{z}-axis and a constant normalized precession rate $\beta$ where time $\tau$ is measured in secular units (same as \cite{klein2023,klein2024}).

The scenario of a constant rate precessing quadrupole potential acting on a Keplerian orbit of a test particle is of importance since it is equivalent (under some restricting assumptions) to higher multiplicity systems \citep{hamers2017,petrovich2017} which lead the Keplerian orbit to high eccentricities from a wide range of initial conditions \citep{pejcha2013,hamers2015,hamers2017a,petrovich2017,fang2018,grishin18,liu2019,safarzadeh2020,hamers2020,bub2020,grishin22}. High eccentricity is a necessary ingredient in proposed formation channels of a variety of astrophysical phenomena \citep{naoz2011,katz2011,fabrycky2007,munoz20,oconnor21,stephan21,thompson2011,katz2012,antonini2012,liu2018,Melchor2024,naoz2012,teyssandier2013,petrovich2015,stephan2016,liu2018,Angelo2022,Melchor2024}.

In \citet{klein2023,klein2024} this problem was solved analytically for librating KLCs (a solution that can be refined and implemented to the rotating KLCs as well).
In two previous papers (\citet{Klein2024_pendulum}, hereafter referred to as \citepaperI, and \citet{Klein2024_CDA}, hereafter referred to as \citepaperII), we have shown that in the high eccentricity regime, The Eccentric Kozai Lidov (EKL) \citep{katz2011,naoz2011,lithwick2011,naoz2013} with and without Brown's Hamiltonian \citep{brown1936a,brown1936b,brown1936c,Soderhjelm1975,Cuk2004,Breiter2015,luo2016,Will2021,tremaine2023} can be described using a simple pendulum model allowing a derivation of an explicit flip criterion.

In this Letter, we simplify the approach of \citet{klein2023,klein2024} to analytically solve the effect of the precession of an outer quadrupole potential for precession rates in the vicinity of the frequency of the KLCs. We show that this resonant combined effect of proximity of frequencies between the KLCs and the perturbing potential can be described also by a simple pendulum. We also show that regular KLCs at high eccentricity are described using a simple pendulum allowing an analytical derivation of the frequencies for comparison with the precession rate of the perturbing potential. Following the approach of \citepaperI and \citepaperII we use the analytic expressions of the regular KLC simple pendulum model to construct the analytic solution when the potential precesses. We compare the analytic solution to the numerical results of the double averaged equations. We explore a large phase space of initial conditions but we restrict our analysis to $\alpha\ll1$.

\section{Equations of Motion}
The long-term dynamics of the test particle can be parameterized
by two dimensionless orthogonal vectors $\mathbf{j}=\mathbf{J}/\sqrt{GMa}$, where $\mathbf{J}$ is the specific angular momentum vector, and
$\mathbf{e}$ a vector pointing in the direction of the pericenter
with magnitude $e$. In the secular approximation, where the equations are averaged over the orbit, $a$ is constant with time while $\mathbf{j}$ and
$\mathbf{e}$ evolve (after double averaging) according to (same as Equations 3 in \citet{klein2024})
\begin{eqnarray}
\frac{d\mathbf{j}}{d\tau}=&\frac{3}{4}\left(\left(\mathbf{j}\cdot\mathbf{\hat{j}}_{\text{outer}}\right)\mathbf{j}-5\left(\mathbf{e}\cdot\mathbf{\hat{j}}_{\text{outer}}\right)\mathbf{e}\right)\times\mathbf{\hat{j}}_{\text{outer}} \cr
\frac{d\mathbf{e}}{d\tau}=&\frac{3}{2}\left(\mathbf{j}\times\mathbf{e}\right)-\frac{3}{4}\left(5\left(\mathbf{e}\cdot\mathbf{\hat{j}}_{\text{outer}}\right)\mathbf{j}-\left(\mathbf{j}\cdot\mathbf{\hat{j}}_{\text{outer}}\right)\mathbf{e}\right)\times\mathbf{\hat{j}}_{\text{outer}}. \cr
 \label{eq:secular_equations}
\end{eqnarray}

\section{Pure KLCs}\label{sec:Pure_KLCs}
When the perturbing potential is not precessing the periodic Kozai-Lidov cycles (KLCs) solved analytically by \citet{kozai62,lidov1962} are obtained (for a recent review see \citet{naoz2016})\footnote{See recent historical overview including earlier relevant work by \citet{vonZeipel1910} in \citet{ito_2019}.}. In that case, the external potential is constant in time and axis-symmetric - thus having a constant of motion $\mathbf{j}\cdot\mathbf{\hat{j}}_{\text{outer}}=j_z$. A second constant emerges from the double averaged potential and $\mathbf{j}\cdot\mathbf{\hat{j}}_{\text{outer}}$,
\begin{equation}\label{eq:CK}
C_{K}\equiv e^{2}-\frac{5}{2}\left(\mathbf{e}\cdot\mathbf{\hat{j}}_{\text{outer}}\right)^{2}=e^2-\frac{5}{2}e^2_z.
\end{equation}
The sign of the constant $C_K$ parts the phase space to two classes aligned with whether the argument of pericenter of the Keplerian orbit librates around $\frac{\pi}{2}$ or $-\frac{\pi}{2}$ (\textit{librating} cycles) when $C_K<0$, or if it goes through all values $\left[0,2\pi\right]$ when $C_K>0$ (\textit{rotating} cycles).

We are interested in the occurrence of high eccentricities so we focus on the range of $\left|j_z\right|\ll1$.

\subsection{KLC Frequency at $j_z=0$}\label{subsec:KLC_frequency_at_jz_0}
In the case of $j_z=0$ the vector $\mathbf{j}$ oscillates back and forth on a straight line in the $\textit{x-y}$ plane crossing the origin, so $e$ achieves a maximal value twice in this oscillation. At this limit there is no rotation of this line so there is no longitudinal precession. Therefore, in this limit, the relevant KLC frequency is the frequency of this oscillation with a period double that of the oscillation of $e$. In this case we can align the {\em{y}}-axis on this straight line where the vector $\mathbf{j}$ oscillates so at all times $j_x=j_z=0$. Since $\mathbf{e}\cdot\mathbf{j}=0$ we have the vector $\mathbf{e}$ inside the {\em{x-z}} plane, i.e $e_y=0$ and $C_K=e^2_x-\frac{3}{2}e^2_z$. At this coordinate system, $\mathbf{\hat{j}}_{\text{outer}}$ has an initial phase in the {\em{x-y}} plane, $\Omega^0_{\mathbf{\hat{j}}_{\text{outer}}}$.

As detailed in appendix \ref{app:pure_Kozai_jz_0_simple_pendulum} - for regular KLCs at $j_z=0$ the equations of motion are equivalent to a simple pendulum with a velocity $\propto e_x$ and libration or rotation of the pendulum corresponds to librating or rotating KLCs. The frequency of the pendulum is defined by 
\begin{equation}\label{eq:omega0}
    \omega_0=\frac{2\pi}{T}
\end{equation}
with $T$ a function of $C_K$ alone through Equations \ref{eq:period_rotating_pendulum} and \ref{eq:period_librating_pendulum}
\begin{equation}
\begin{aligned}
    T&=\frac{8}{3}\sqrt{\frac{2}{2C_{K}+3}}K\left(x\right) \quad & C_K>0 \\ 
    T&=\frac{8}{3}\sqrt{\frac{2}{3\left(1-C_{K}\right)}}K\left(\frac{1}{x}\right) \quad & C_K<0,
\end{aligned}
\label{eq:Period_KLCs_at_jz_0}
\end{equation}
with 
\begin{equation}\label{eq:x}
    x=3\frac{1-C_{K}}{3+2C_{K}}
\end{equation}
and $K\left(m\right)$ the complete elliptic function of the first kind. These expressions for $T$ are consistent with numerically integrating Equation 27 of \citet{antognini2015} at $j_z=0$ (Note that the period in \citet{antognini2015} is for the eccentricity $e$ and thus is half the period of the oscillation of $\mathbf{j}$ on the straight line). Since the behavior of the KLC is different depending on the sign of $C_K$ we restrict our analysis to numerical results such that $C_K$ did not cross zero during the evolution.

\section{Resonating frequencies}\label{sec:resonting_frequencies}

\begin{figure}\label{fig:top_omega0_bottom_numeical_deltajz_vs_Ck0_beta_1.8_alpha_1deg}
 \begin{centering}
    \begin{centering}
 \includegraphics[scale=0.45]{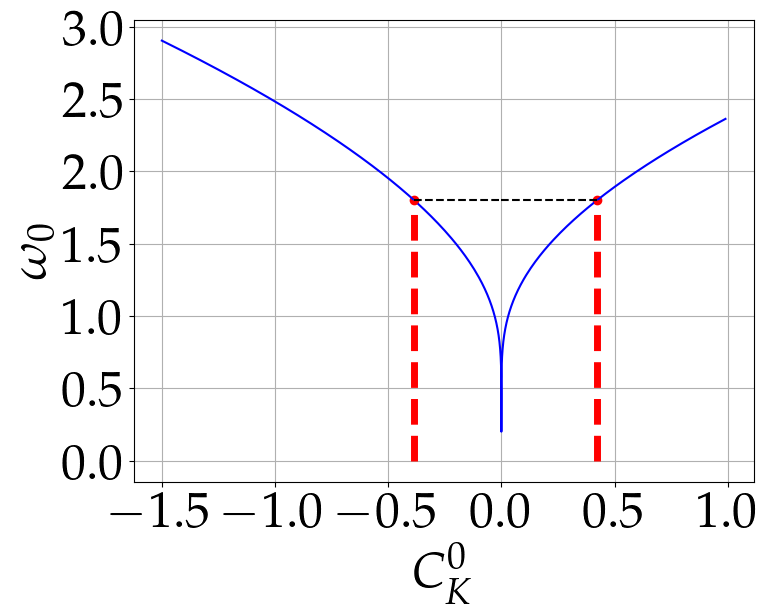}
 \par\end{centering}
 \includegraphics[scale=0.45]{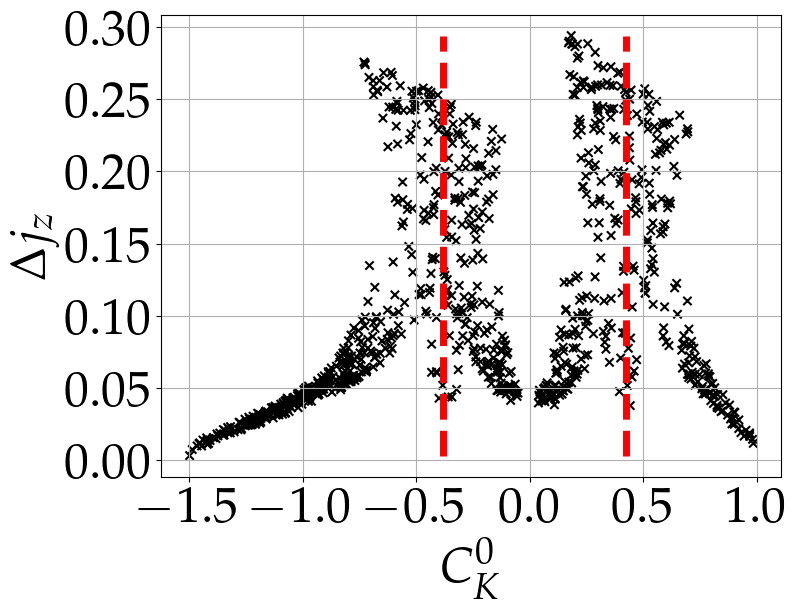}
  \par\end{centering}
 \caption{Top panel: $\omega_0$ (Equations \ref{eq:omega0}-\ref{eq:x}) vs. $C_K$. A vertical black line denotes the value $\beta=1.8$ and red dashed horizontal lines denote the values $C_K\approx-0.38$ and $C_K\approx0.42$ where $\omega_0=\beta=1.8$ (through Equations \ref{eq:omega0}-\ref{eq:x}). Bottom panel: $\Delta j_z$ vs. $C^0_K$ from numerical integration (up to $\tau=300$) of Equations \ref{eq:jOuter_as_a_function_of_tau}-\ref{eq:secular_equations} with $\alpha=1^\circ$ and  $\beta=1.8$ for randomly selected initial conditions with $j^0_z=0$ (uniformly distributed in $e_x,j_y,e_z$) . The red dashed horizontal lines are the same as the top panel.
 \label{fig:jz_amplitude_with_horizontal_Ck0}}
\end{figure}

The quadrupole potential in the problem we solve is precessing at a constant angular frequency $\beta$. As presented in \citet{klein2023}, the prominent influence of the precession is obtained when there is a resonance between the frequency of the precession of the perturbing potential and the relevant frequency of the KLC oscillations, $\omega_0$ (Equations \ref{eq:omega0}-\ref{eq:x}). An example of the frequencies correspondences in shown in Figure \ref{fig:jz_amplitude_with_horizontal_Ck0}. In the top panel, the function $\omega_0$ (Equations \ref{eq:omega0}-\ref{eq:x}) is plotted versus $C_K$. A black dashed vertical line is shown for a value of $\beta=1.8$ and red dashed horizontal lines mark the values of $C_K$ where $\omega_0=\beta$ and resonance is expected ($C_K\approx-0.38$ and $C_K\approx0.42$). In the bottom panel, the amplitude of change in $j_z$ throughout the numerical integration of Equations \ref{eq:jOuter_as_a_function_of_tau}-\ref{eq:secular_equations} ($\Delta j_z=j^{\text{max}}_z-j^{\text{min}}_z$, up to $\tau=300$) versus $C^0_K$ is shown for $\alpha=1^\circ$, $\beta=1.8$ and randomly selected initial conditions with $j^0_z=0$ (uniformly distributed in $e_x,j_y,e_z$) using black crosses. The red dashed horizontal lines are the same as in the top panel, i.e at the expected values of $C_K$ to obtain the most significant effect of the precessing perturbance.

\section{Averaged Equations}\label{sec:averaged_equations}
\begin{figure}
 \begin{centering}
 \includegraphics[scale=0.45]{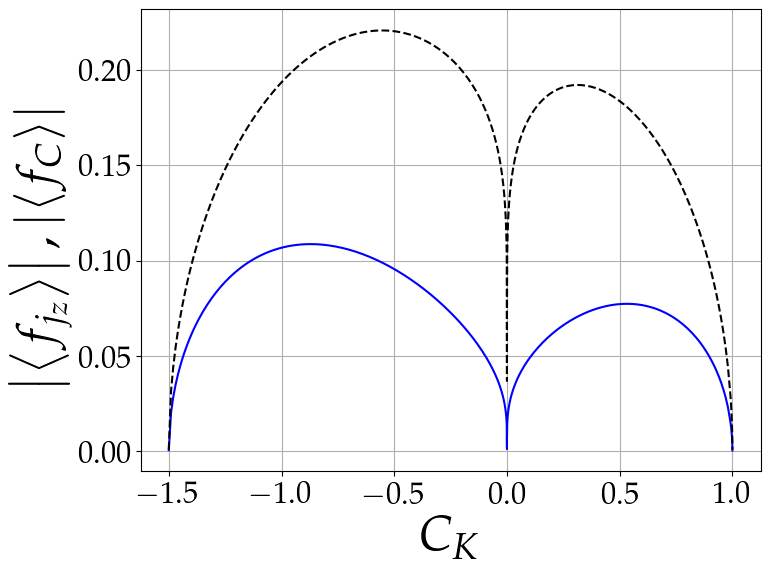}
  \par\end{centering}
 \caption{$\left|\left\langle f_{C} \right\rangle\right|$ (solid blue) and $\left|\left\langle f_{{j}_{z}} \right\rangle\right|$ (dashed black) as function of $C_K$ (see Equations \ref{eq:fjz_fck_definition}).
 \label{fig:numerically_evaluated_averages}}
\end{figure}

In the problem we solve, the perturbing potential is precessing and $j_z$ and $C_K$ are no longer constants but rather slowly evolve. The slow change of $j_z$ and $C_K$ at the vicinity of $\left|j_z\right| \ll 1$ follows Equations \ref{eq:jOuter_as_a_function_of_tau}-\ref{eq:secular_equations} which at the coordinate system where $j_x=e_y=j_z=0$ and $C_K=e^2_x-\frac{3}{2}e^2_z$, up to first order in $\alpha$ become
\begin{equation}
\begin{aligned}
    \dot{j}_z & =-\alpha\frac{15}{4} e_{z}e_{x}\sin\left(\Omega^0_{\mathbf{\hat{j}}_{\text{outer}}}-\beta\tau\right) \\
    \dot{C}_K & = -\alpha\frac{45}{2}j_{y}\left(\frac{1}{3}e_{x}^{2}+\frac{1}{2}e_{z}^{2}\right)\cos\left(\Omega^0_{\mathbf{\hat{j}}_{\text{outer}}}-\beta\tau\right).
\end{aligned}
\end{equation}
At the vicinity of $C_K$ values where $\omega_0\left(C_K\right)\approx \beta$ - since $C_K$ is slowly changing, the phase difference between the oscillatory KLC phenomena and the constant precession of the perturbing potential 
\begin{equation}\label{eq:phi}
    \phi=\left(\omega_0 - \beta\right) \tau + \Omega^0_{\mathbf{\hat{j}}_{\text{outer}}}
\end{equation}
is also slowly changing. We next average the equations of the slow variables over KLCs with $j_z=0$ assuming $\phi$ is constant during a KLC.  %obtaining 
% \begin{equation}\label{eq:averaged_jzdot_Ckdot_full}
% \begin{aligned}
%     \left\langle \dot{j}_{z}\right\rangle & =-\alpha\frac{15}{4}\left(\begin{array}{c}
% \left\langle e_{z}e_{x}\cos\omega_{0}\tau\right\rangle \sin\phi\\
% -\left\langle e_{z}e_{x}\sin\omega_{0}\tau\right\rangle \cos\phi
% \end{array}\right) \\
%    \left\langle \dot{C}_{K}\right\rangle & =-\alpha\frac{45}{2}\left(\begin{array}{c}
% \left\langle j_{y}\left(\frac{1}{3}e_{x}^{2}+\frac{1}{2}e_{z}^{2}\right)\sin\omega_{0}\tau\right\rangle \sin\phi+\\
% \left\langle j_{y}\left(\frac{1}{3}e_{x}^{2}+\frac{1}{2}e_{z}^{2}\right)\cos\omega_{0}\tau\right\rangle \cos\phi
% \end{array}\right).
% \end{aligned}
% \end{equation}
Since at each half of a KLC cycle $j_y$ oscillates symmetrically from zero to $j^\text{max}$ (or $j^\text{min}$) but the $\cos$ term oscillates symmetrically around zero we obtain 
\begin{equation}
    \left\langle j_{y}\left(\frac{1}{3}e_{x}^{2}+\frac{1}{2}e_{z}^{2}\right)\cos\omega_{0}\tau\right\rangle=0.
\end{equation}
Similarly, since at each half of a KLC $e_ze_x$ oscillates symmetrically around zero ($e_z$ for a rotating KLC and $e_x$ for a librating KLC) we obtain
\begin{equation}
    \left\langle e_{z}e_{x}\sin\omega_{0}\tau\right\rangle=0.
\end{equation}
All in all we have
\begin{equation}\label{eq:averaged_jzdot_Ckdot}
\begin{aligned}
    \left\langle \dot{j}_{z}\right\rangle & =-\alpha\frac{15}{4}
\left\langle f_{j_z} \right\rangle \sin\phi
 \\
   \left\langle \dot{C}_{K}\right\rangle & =-\alpha\frac{45}{2}
\left\langle f_{C} \right\rangle \sin\phi
.
\end{aligned}
\end{equation}
with 
\begin{equation}\label{eq:fjz_fck_definition}
\begin{aligned}
    f_{j_z} & = e_{z}e_{x}\cos\omega_{0}\tau
 \\
    f_{C} & = j_{y}\left(\frac{1}{3}e_{x}^{2}+\frac{1}{2}e_{z}^{2}\right)\sin\omega_{0}\tau
.
\end{aligned}
\end{equation}
The averages $\left\langle f_{j_z} \right\rangle$,$\left\langle f_{C} \right\rangle$ are made over KLCs at $j_z=0$ and are independent of $\beta$ so can be calculated numerically as functions of $C_K$ alone. The sign of the resulting average is the same as the sign of $e_z e_x$ of the initial conditions. The absolute value of the averages are plotted in Figure \ref{fig:numerically_evaluated_averages}. 

\section{Simple Pendulum}\label{sec:simple_pendulum}
The evolution of $\phi$ (Equation \ref{eq:phi}) is obtained as follows. Denote the following variable  (with $\dot{\omega}_0/\omega_0 \ll 1$)
\begin{equation}\label{eq:delta}
    \delta = \dot{\phi} = \omega_0 - \beta
\end{equation}
with its time derivative
% \begin{equation}\label{eq:deltadot}
%     \dot{\delta} = \frac{d\omega_{0}}{dC_{K}} \left\langle \dot{C}_{K}\right\rangle = \frac{d\omega_{0}}{dC_{K}}\left(-\alpha\frac{45}{2}\left\langle j_{y}\left(\frac{1}{3}e_{x}^{2}+\frac{1}{2}e_{z}^{2}\right)\sin\omega_{0}\tau\right\rangle \right)\sin\phi.
% \end{equation}
\begin{equation}\label{eq:deltadot}
    \dot{\delta} = \frac{d\omega_{0}}{dC_{K}} \left\langle \dot{C}_{K}\right\rangle = -\alpha\frac{45}{2}\left\langle f_{C}\right\rangle \frac{d\omega_{0}}{dC_{K}} \sin\phi .
\end{equation}
 with $d\omega_0/dC_K$ given analytically by (through Equation \ref{eq:x}):

for $C_K>0$
\begin{equation}
    \frac{d\omega_{0}}{dC_{K}}=\pi\frac{3}{8}\frac{1}{K^{2}\left(x\right)}\sqrt{\frac{2}{3+2C_{K}}}\left(K\left(x\right)+\frac{15}{3+2C_{K}}\frac{E\left(x\right)-\left(1-x\right)K\left(x\right)}{2\left(1-x\right)x}\right)
\label{eq:domega0_dcK_for_rotating_KLCs_at_jz_0}    
\end{equation}
and for $C_K<0$    
\begin{equation}
    \frac{d\omega_{0}}{dC_{K}}=-\pi\frac{9}{8}\frac{1}{K^{2}\left(\frac{1}{x}\right)}\sqrt{\frac{2}{3\left(1-C_{K}\right)}}\left(\begin{array}{c}
\frac{1}{2}K\left(\frac{1}{x}\right)\\
+\frac{5}{3\left(1-C_{K}\right)}\frac{E\left(\frac{1}{x}\right)-\left(1-\frac{1}{x}\right)K\left(\frac{1}{x}\right)}{2\left(1-\frac{1}{x}\right)\frac{1}{x}}
\end{array}\right)
\label{eq:domega0_dcK_for_librating_KLCs_at_jz_0}
\end{equation}
with $E\left(m\right)$ the complete elliptic function of the second kind.

Since $\alpha \ll 1$ and our focus is on cases where $C_K$ keeps its sign, i.e $C_K$ does not change much with respect to $C^0_K$. Similar to \citepaperI and \citepaperII we use that fact to evaluate the different functions of $C_K$ at $C^0_K$ making them all constant values so Equations \ref{eq:delta}-\ref{eq:deltadot} become those of a simple pendulum with angle $\phi$ for $\left\langle f_{C}\right\rangle \frac{d\omega_{0}}{dC_{K}}>0$ and $\phi+\pi$ otherwise where as $\delta$ is the velocity of the pendulum. Using Equations \ref{eq:averaged_jzdot_Ckdot} and \ref{eq:deltadot} the following holds 
\begin{equation}\label{eq:deltadot_jzdot}
\dot{\delta}=6\frac{\left\langle f_{C}\right\rangle }{\left\langle f_{j_{z}}\right\rangle }\frac{d\omega_{0}}{dC_{K}}\left\langle \dot{j}_{z}\right\rangle
\end{equation}
so up to a constant (depending on initial conditions $j^0_z,C^0_K$) the velocity of the pendulum determines the evolution of $\left\langle {j}_{z}\right\rangle$.

\subsection{Precession of $\mathbf{e}$ for rotating KLCs}
For rotating KLCs, slow precession of the vector $\mathbf{e}$ in the \textit{x-y} plane when $j_z$ does deviate from zero can be comparable to the effect of the precession of the perturbing potential and therefore should be taken into account. This precession induces a slight change in the coordinate system we use assuming the vector $\mathbf{e}$ is not precessing in the \textit{x-y} plane (which is relevant when $j_z$ is strictly zero). This correction can incorporated through $\phi$ and $\delta$. Using the variables $i_e$ and $\Omega_e$ as in \cite{katz2011} the eccentricity vector $\mathbf{e}$ is given by 
\begin{equation}
  \mathbf{e}=e\left(\sin i_e \cos \Omega_e,\sin i_e \sin\Omega_e,\cos i_e\right)
\end{equation} 
and the slow evolution of $\Omega_e$ (the slow precession of $\mathbf{e}$ in the \textit{x-y} plane) up to the leading order in $j_z$ is given by \citep{katz2011} 
\begin{equation}\label{eq:omega_e_dot}
    \dot{\Omega}_e = \left<f_\Omega\right> j_z
\end{equation}
where (using Equation \ref{eq:x})
\begin{equation}\label{eq:fOmega}
  \left\langle f_{\Omega}\right\rangle =\frac{6E\left(x\right)-3K\left(x\right)}{4K\left(x\right)}.
\end{equation}
Noting that in the coordinate system we use (where $\mathbf{e}$ is not precessing in the \textit{x-y} plane and is aligned in it on the \textit{x} axis), $\Omega_e=0$, so Equation \ref{eq:phi} is equivalent to defining 
\begin{equation}
    \phi=\left(\omega_0 - \beta\right) \tau + \left(\Omega^0_{\mathbf{\hat{j}}_{\text{outer}}} - \Omega_e \right)
\end{equation}
leading to 
\begin{equation}
    \delta = \dot{\phi} = \omega_0 - \beta -\dot{\Omega}_e
\end{equation}
and using Equation \ref{eq:omega_e_dot} the equations of $\phi$ and $\delta$ retain the simple pendulum structure with approximating $\left\langle f_{\Omega}\right\rangle,\left\langle f_{C}\right\rangle,\left\langle f_{j_Z}\right\rangle,\frac{d\omega_{0}}{dC_{K}}$ as constants evaluated at $C^0_K$
\begin{equation}\label{eq:phi_dot_delta_dot_rotating}
\begin{aligned}
\dot{\phi} & = \delta\\
    \dot{\delta} & = -\alpha \left(1-\left\langle f_{\Omega}\right\rangle \left\langle f_{j_{z}}\right\rangle \frac{1}{6}\frac{1}{\left\langle f_{C}\right\rangle }\frac{1}{\frac{d\omega_{0}}{dC_{K}}}\right) \frac{45}{2}\left\langle f_{C}\right\rangle \frac{d\omega_{0}}{dC_{K}} \sin\phi.
    \end{aligned}
\end{equation}
Using Equation \ref{eq:averaged_jzdot_Ckdot} the connection between $\delta$ and $j_z$ is also slightly corrected with
\begin{equation}\label{eq:deltadot_jzdot_fOmega}
\dot{\delta}=\left(1-\left\langle f_{\Omega}\right\rangle \left\langle f_{j_{z}}\right\rangle \frac{1}{6}\frac{1}{\left\langle f_{C}\right\rangle }\frac{1}{\frac{d\omega_{0}}{dC_{K}}}\right) 6  \frac{\left\langle f_{C}\right\rangle}{\left\langle f_{j_z} \right\rangle} \frac{d\omega_{0}}{dC_{K}} \left\langle \dot{j}_{z}\right\rangle
\end{equation}

\begin{figure}
 \begin{centering}
 \includegraphics[scale=0.4]{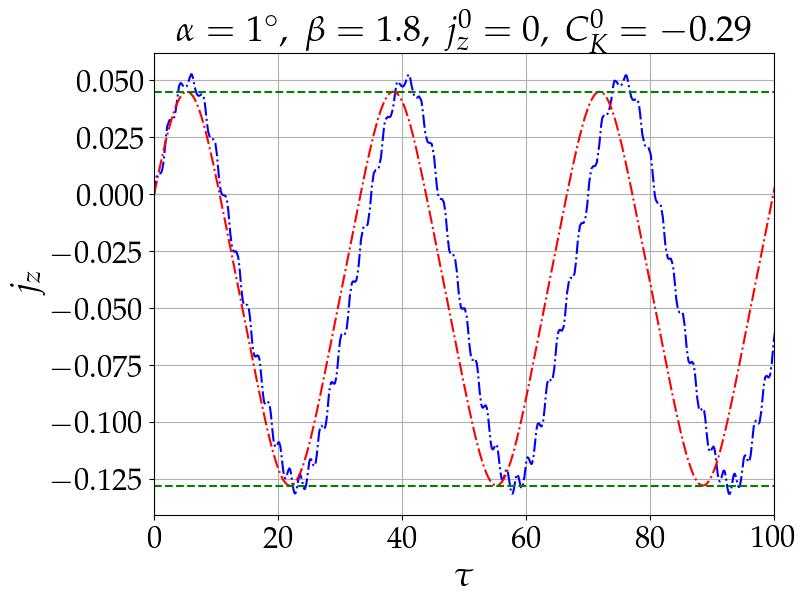}
 \par\end{centering}
 \begin{centering}
 \includegraphics[scale=0.4]{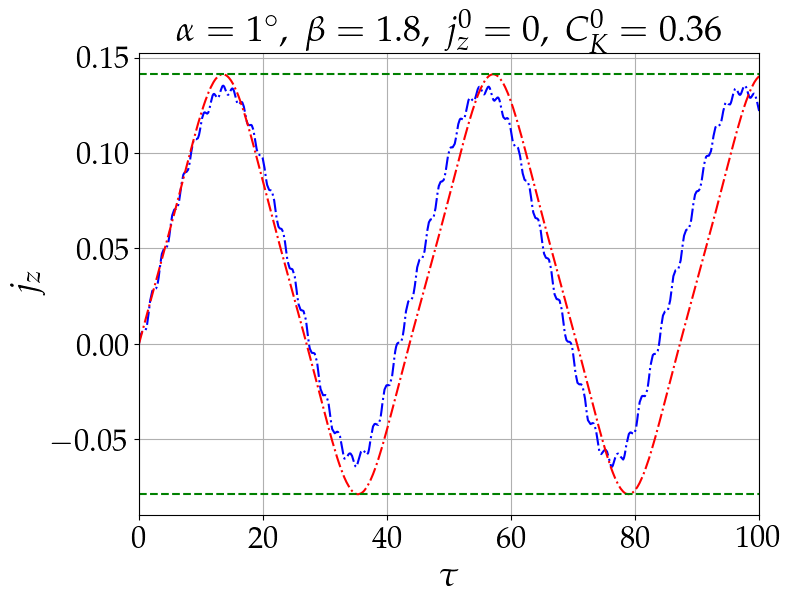}
 \par\end{centering}
 % \begin{centering}
 % \includegraphics[scale=0.25]{epsilonoct_0.001_jz_vs_Omega_e_up_to_tau_3161.84_no_box.png}
 % \par\end{centering}
 \caption{Results of a numeric integration of the double averaged equations (blue) (Equations \ref{eq:secular_equations}) along with the result of a simple pendulum (red, Equations \ref{eq:delta}-\ref{eq:deltadot},\ref{eq:deltadot_jzdot} in the top panel and Equations \ref{eq:phi_dot_delta_dot_rotating},\ref{eq:deltadot_jzdot_fOmega}, bottom panel). Presented are two examples, librating KLC (top panel) and rotating KLC (bottom panel). The values of the initial conditions are shown above each plot. Shown is $j_z$ as a function of (normalized) time. The two green horizontal lines are the maximum and minimum values of $j_z$ of the simple pendulum.
 \label{fig:jz_vs_tau_example}}
\end{figure}

Two examples of a numerical integration of Equations \ref{eq:secular_equations} (for rotating and librating KLCs) compared with the solution of Equations \ref{eq:delta}-\ref{eq:deltadot} (top panel) with the approximation of constant $\left\langle f_{C}\right\rangle$, and using Equation \ref{eq:deltadot_jzdot} with constant $\left\langle f_{j}\right\rangle,\left\langle f_{C}\right\rangle$ (all evaluated at $C^0_K$) to obtain $j_z$ is shown in Figure \ref{fig:jz_vs_tau_example} (top panel) and Equations \ref{eq:phi_dot_delta_dot_rotating} with Equation \ref{eq:deltadot_jzdot_fOmega} (bottom panel). As can be seen - for the examples shown - the long term evolution of $j_z$ is successfully approximated by equations of a simple pendulum.

\section{Maximal Deviation of $j_z$ from $j^0_z=0$}\label{sec:jzmax_from_jz0}
The maximal and minimal values of $j_z$ obtained in $1000$ instances with $j^0_z=j^0_x=e^0_y=0$ and randomly chosen initial conditions (uniformly distributed in $j_y,e_z$ and $\Omega^0_{\mathbf{\hat{j}}_{\text{outer}}}$) are evaluated both numerically through simulating Equations \ref{eq:jOuter_as_a_function_of_tau}-\ref{eq:secular_equations} and analytically with the simple pendulum models. Since the analytic model is a simple pendulum model, obtaining $\delta^{\text{max}}$ and $\delta^{\text{min}}$ is straightforward. $j^{\text{max}}_z$ and $j^{\text{min}}_z$ are obtained through the connection between the velocity of the pendulum, $\delta$, and $j_z$ using Equations \protect\ref{eq:delta}-\protect\ref{eq:deltadot_jzdot} for librating KLCs (with $C^0_K<0$) and Equations \protect\ref{eq:phi_dot_delta_dot_rotating}-\protect\ref{eq:deltadot_jzdot_fOmega} for rotating KLCs (with $C^0_K>0$).

\subsection{$\alpha$ dependence}\label{srbsec:alpha_depencdence}
\begin{figure}
 % \begin{centering}
 % \includegraphics[scale=0.21]{jz_amplitude_vs_initial_CK_pendulum_eps_t_0.01_alpha_deg_0.0625_beta_1.8_jz_0.0_negate_oscillating_False_fOmega_correction_rotating_jz_0_N_997.png}
 %  \includegraphics[scale=0.21]{jz_amplitude_vs_initial_CK_pendulum_eps_t_0.01_alpha_deg_0.125_beta_1.8_jz_0.0_negate_oscillating_False_fOmega_correction_rotating_jz_0_N_995.png}
 %  \par\end{centering}
 %  \begin{centering}
 % \includegraphics[scale=0.21]{jz_amplitude_vs_initial_CK_pendulum_eps_t_0.01_alpha_deg_0.25_beta_1.8_jz_0.0_negate_oscillating_False_fOmega_correction_rotating_jz_0_N_985.png}
 %  \includegraphics[scale=0.21]{jz_amplitude_vs_initial_CK_pendulum_eps_t_0.01_alpha_deg_0.5_beta_1.8_jz_0.0_negate_oscillating_False_fOmega_correction_rotating_jz_0_N_971.png}
 %  \par\end{centering}
 %  \begin{centering}
 % \includegraphics[scale=0.21]{jz_amplitude_vs_initial_CK_pendulum_eps_t_0.01_alpha_deg_1.0_beta_1.8_jz_0.0_negate_oscillating_False_fOmega_correction_rotating_jz_0_N_929.png}
 %  \includegraphics[scale=0.21]{jz_amplitude_vs_initial_CK_pendulum_eps_t_0.01_alpha_deg_2.0_beta_1.8_jz_0.0_negate_oscillating_False_fOmega_correction_rotating_jz_0_N_794.png}
 %  \par\end{centering}
   \begin{centering}
 \includegraphics[scale=0.21]{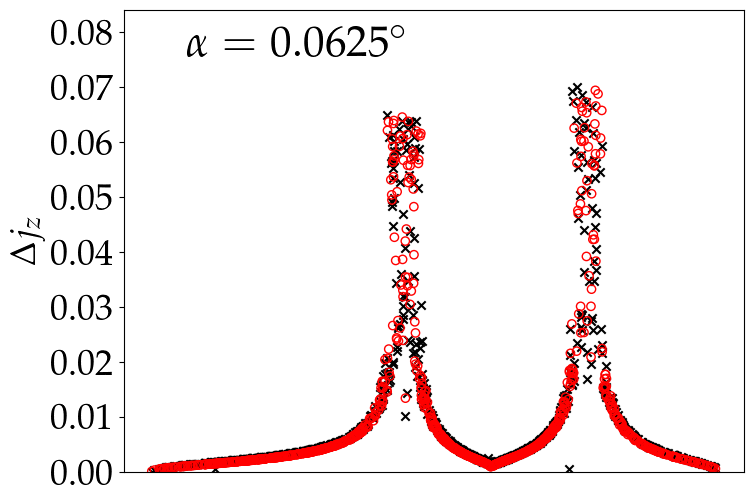}
  \includegraphics[scale=0.21]{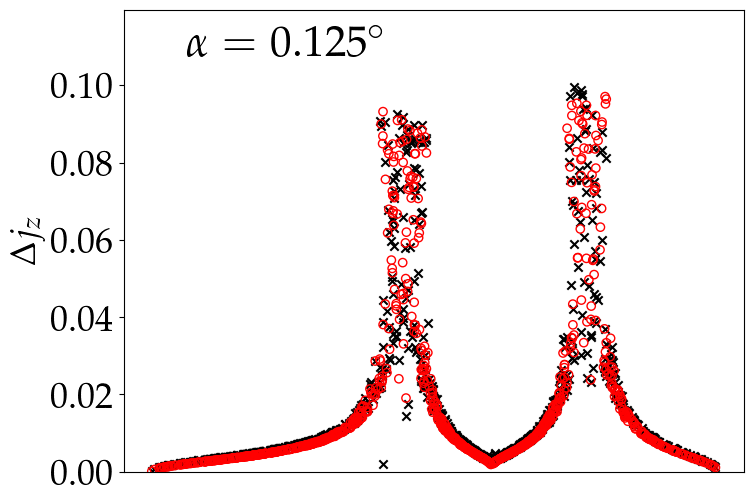}
  \par\end{centering}
\begin{centering}
 \includegraphics[scale=0.21]{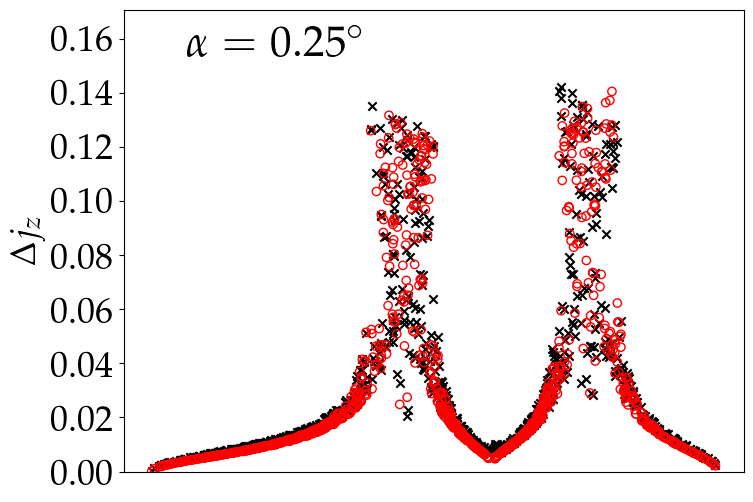}
  \includegraphics[scale=0.21]{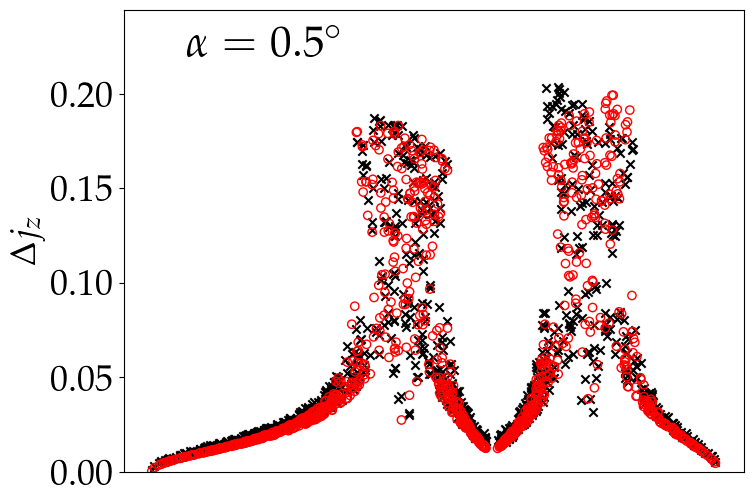}
  \par\end{centering}
\begin{centering}
 \includegraphics[scale=0.21]{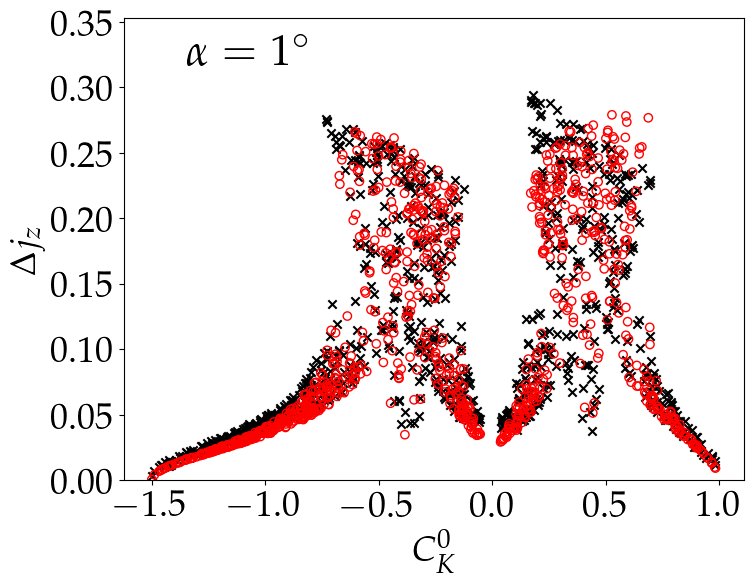}
  \includegraphics[scale=0.21]{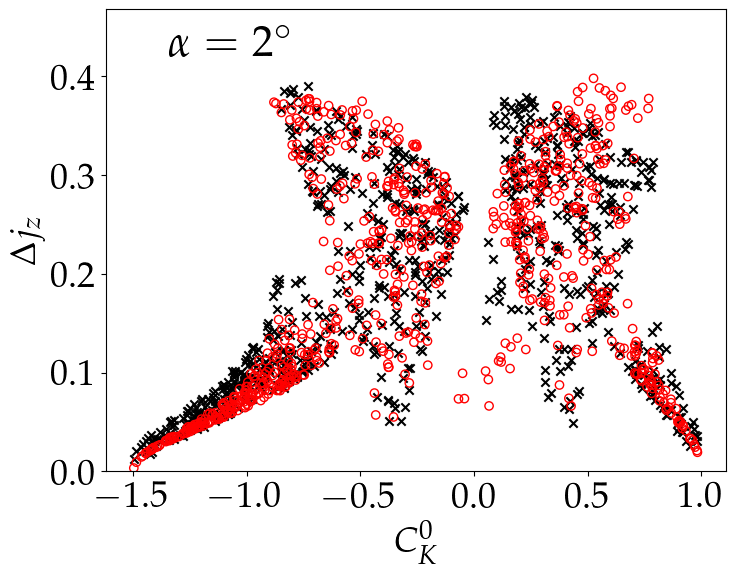}
  \par\end{centering}
\caption{Amplitude of the change in $j_z$ vs. $C^0_K$ for precession rate of $\beta=1.8$ and different values of $\alpha$ (written explicitly inside each subplot). Shown are initial conditions with $j^0_z=j^0_x=e^0_y=0$ and randomly chosen values from a uniformly distributed $j_y,e_z$ and $\Omega^0_{\mathbf{\hat{j}}_{\text{outer}}}$. Shown are results from a full numerical solution of Equations \ref{eq:jOuter_as_a_function_of_tau}-\ref{eq:secular_equations} using black crosses. Red circles denote the prediction of the pendulum model using Equations \protect\ref{eq:delta}-\protect\ref{eq:deltadot_jzdot} for librating KLCs (with $C^0_K<0$) and Equations \protect\ref{eq:phi_dot_delta_dot_rotating}-\protect\ref{eq:deltadot_jzdot_fOmega} for rotating KLCs (with $C^0_K>0$). Note the different $\textit{y}$-axis scale in each subplot.}
 \label{fig:jz_amplitude_vs_CK_constant_beta}
\end{figure}

In Figure \ref{fig:jz_amplitude_vs_CK_constant_beta}, we plot $\Delta j_z$ versus $C^0_K$ for $\beta=1.8$ and various logarithmically spaced values of $\alpha$. Note that the scale of the \textit{y}-axis varies across the panels.
From the analysis of Figure \ref{fig:jz_amplitude_vs_CK_constant_beta}, we observe the following significant features of the pendulum model: (a) As mentioned in section \ref{sec:resonting_frequencies} the location of the high amplitude of change in $j_z$ in the $C^0_K$ phase space corresponds to the value of $C^0_K$ where $\omega_0=\beta$. As can be seen, this location is naturally reconstructed in the simple pendulum model. (b) The amplitude of the effect on $\Delta j_z$ is successfully approximated by the analytic solution. (c) As $\alpha$ increases the width of $C^0_K$ parameters for which $\Delta j_z$ becomes significant increases. This broadening is quantitatively captured by the simple pendulum model. (d) In the vicinity of the increased $\Delta j_z$, as this region widens with increasing $\alpha$, a negative slope of the maximal $\Delta j_z$ with respect to $C^0_K$ emerges. For librating KLCs ($C^0_K<0$), this trend is successfully reconstructed by the simple pendulum model. However, for rotating KLCs ($C^0_K>0$), while the model accurately approximates the maximal value, it does not reproduce the negative slope.   

\subsection{$\beta$ dependence}\label{srbsec:beta_depencdence}
\begin{figure}
 % \begin{centering}
 % \includegraphics[scale=0.45]{jz_amplitude_vs_initial_CK_pendulum_eps_t_0.01_alpha_deg_1.0_beta_1.5_jz_0.0_negate_oscillating_False_fOmega_correction_rotating_jz_0_N_911.png}
 %  \par\end{centering}
 %  \begin{centering}
 % \includegraphics[scale=0.45]{jz_amplitude_vs_initial_CK_pendulum_eps_t_0.01_alpha_deg_1.0_beta_1.8_jz_0.0_negate_oscillating_False_fOmega_correction_rotating_jz_0_N_929.png}
 %  \par\end{centering}
 %  \begin{centering}
 % \includegraphics[scale=0.45]{jz_amplitude_vs_initial_CK_pendulum_eps_t_0.01_alpha_deg_1.0_beta_2.1_jz_0.0_negate_oscillating_False_fOmega_correction_rotating_jz_0_N_932.png}
 %  \par\end{centering}
\begin{centering}
 \includegraphics[scale=0.21]{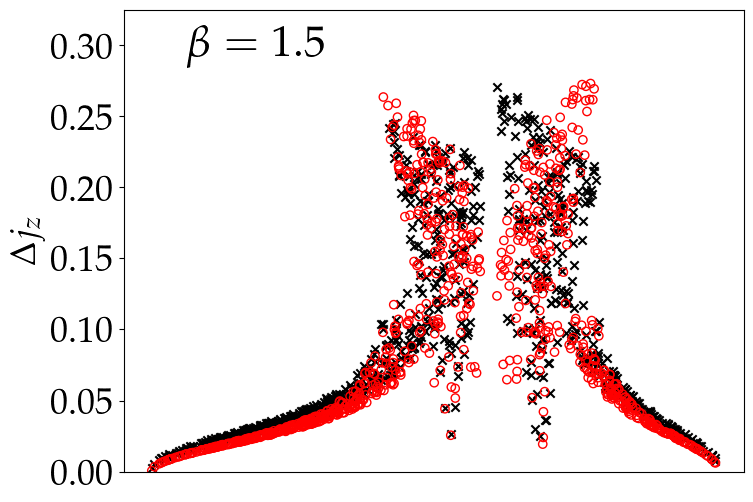}
  \includegraphics[scale=0.21]{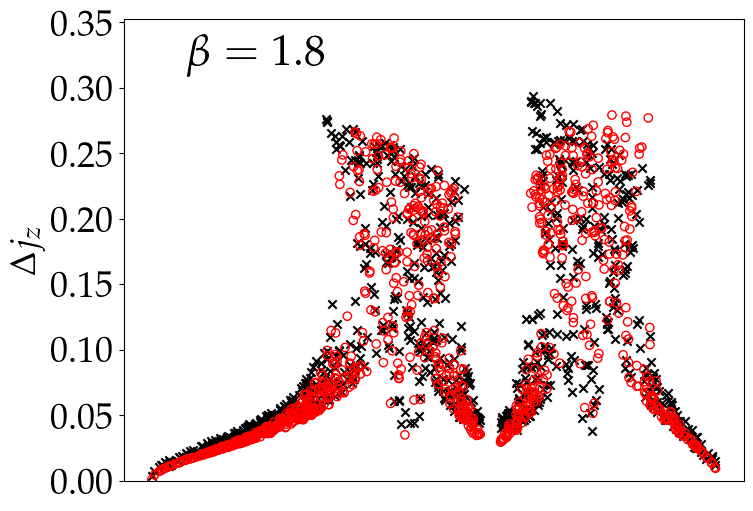}
  \par\end{centering}
\begin{centering}
 \includegraphics[scale=0.21]{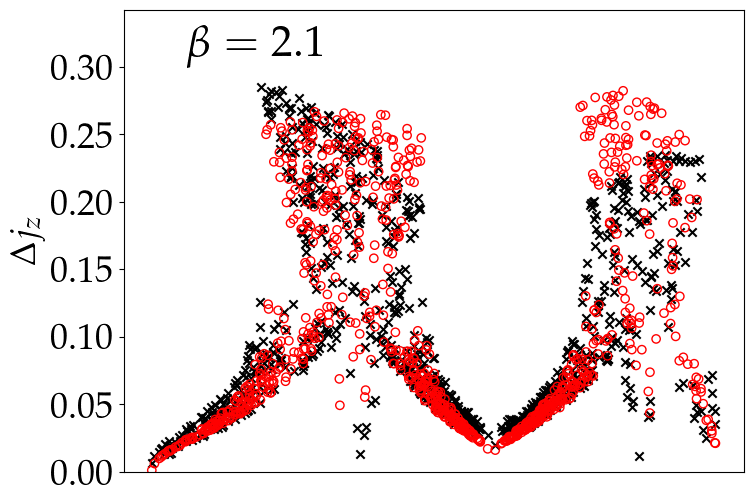}
  \includegraphics[scale=0.21]{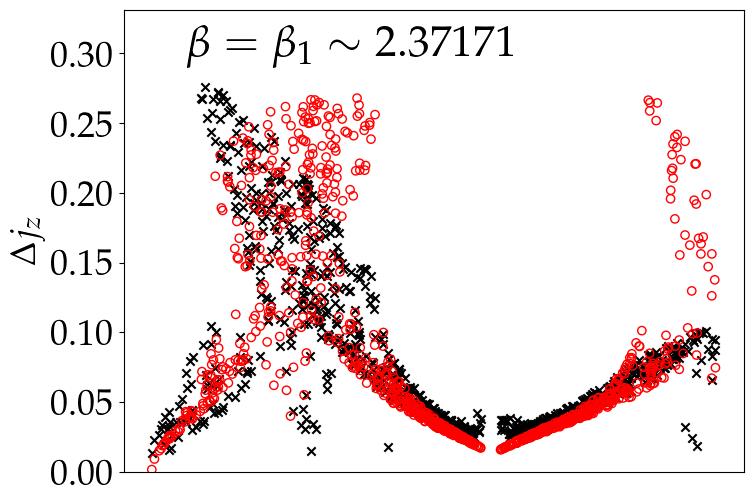}
  \par\end{centering}
\begin{centering}
 \includegraphics[scale=0.21]{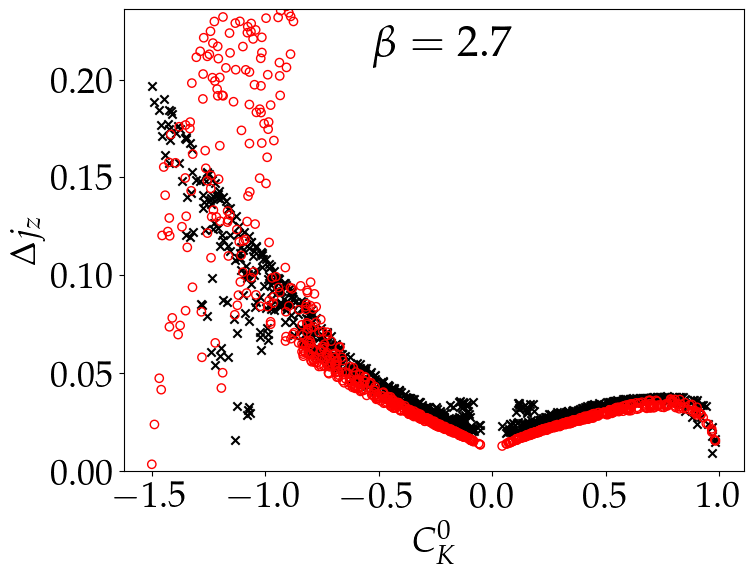}
  \includegraphics[scale=0.21]{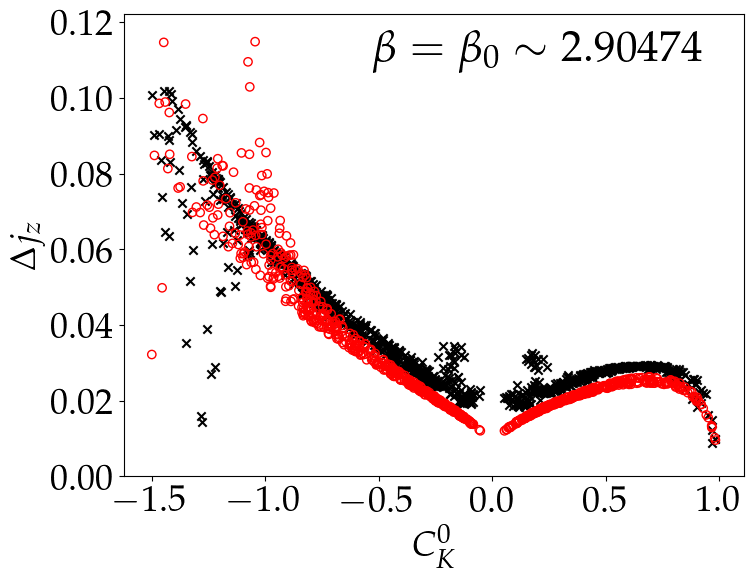}
  \par\end{centering}

\caption{Amplitude of the change in $j_z$ vs. $C^0_K$ for inclination of the perturbing potential of $\alpha=1^\circ$ and different values of $\beta$ (written explicitly inside each subplot). Shown are initial conditions with $j^0_z=j^0_x=e^0_y=0$ and randomly chosen values from a uniformly distributed $j_y,e_z$ and $\Omega^0_{\mathbf{\hat{j}}_{\text{outer}}}$. Shown are results from a full numerical solution of Equations \ref{eq:jOuter_as_a_function_of_tau}-\ref{eq:secular_equations} using black crosses. Red circles denote the prediction of the pendulum model using Equations \protect\ref{eq:delta}-\protect\ref{eq:deltadot_jzdot} for librating KLCs (with $C^0_K<0$) and Equations \protect\ref{eq:phi_dot_delta_dot_rotating}-\protect\ref{eq:deltadot_jzdot_fOmega} for rotating KLCs (with $C^0_K>0$). Note the different $\textit{y}$-axis scale in each subplot.}
 \label{fig:jz_amplitude_vs_CK_constant_alpha}
\end{figure}

Regarding the precession rate of the perturbing potential, as mentioned in section \ref{sec:averaged_equations}, the simple pendulum analysis is centered on precession rates near resonance. Specifically, it focuses on $\beta$ values for which there exists a significant portion of the $C^0_K$ phase space satisfying $\omega_0\left(C_K\right) \approx \beta$ with $C_K$ maintaining a consistent sign during evolution. The critical $\beta$ values at the edges of the $C^0_K$ phase space are defined as:
\begin{equation}\label{eq:beta0}
    \beta_0 = \omega_0\left(C_K=-1.5\right) = \sqrt{\frac{135}{16}}\approx2.905
\end{equation}
as per \citet{klein2023,klein2024} and 
\begin{equation}\label{eq:beta1}
    \beta_1 = \omega_0\left(C_K=1\right) = \sqrt{\frac{45}{8}}\approx2.37.
\end{equation}
From Figure \ref{fig:numerically_evaluated_averages} it is evident that the averages $\left\langle f_{j_z} \right\rangle$,$\left\langle f_{C} \right\rangle$ equal zero at $C_K=-1.5$ and $C_K=1$. Consequently, the relationship between $\delta$ and $j_z$ (as described by Equations \ref{eq:deltadot_jzdot} and \ref{eq:deltadot_jzdot_fOmega}) diverges, and the prediction of the model to the amplitude of change in $j_z$ cannot be fully realized. In Figure \ref{fig:jz_amplitude_vs_CK_constant_alpha} $\Delta j_z$ is plotted against $C^0_K$ for $\alpha=1^\circ$ and various values of $\beta$ with differing \textit{y}-axis scales across panels. The analysis of Figure \ref{fig:jz_amplitude_vs_CK_constant_alpha} reveals the following: (a) As mentioned in section \ref{sec:resonting_frequencies}, as $\beta$ varies, the location of the high amplitude of change in $j_z$ within the $C^0_K$ phase space aligns with the $C^0_K$ value where $\omega_0=\beta$. This dependence is successfully reproduced by the simple pendulum model for $\beta$ values that achieve resonance. (b) As mentioned above, as the value of $C^0_K$ for which $\omega_0=\beta$ approaches the edges of $C_K$ phase space (i.e $-1.5$ and $1$) the prediction of the model diverges. This behavior is visible, for instance, in the panel of $\beta=\beta_1$ at $C^0_K\sim1$ and for $\beta=2.7$ and $\beta=\beta_0$ at the vicinity of $C^0_K=-1.5$. (c) For $\beta=\beta_1$, the resonance for rotating KLCs is obtained at $C^0_K=1$, i.e at the edge, but for librating KLCs there is still a portion of phase space where resonance is obtained and the structure of the numerical results is approximately obtained by the model. (d) Panels for $\beta = 2.7$ and $\beta = \beta_0$ show localized abrupt increases in the numerical results for $\Delta j_z$, arising from higher-order resonances at $C^0_K$ values where $\omega_0 = \frac{1}{2}\beta$. While such abrupt increases were noted for librating KLCs in \citet{klein2024}, their connection to second-order resonance ($\omega_0 = \frac{1}{2}\beta$) was not explicitly mentioned. For rotating KLCs, no first-order resonance occurs (no $C^0_K$ value satisfy $\omega_0 = \beta$), but the overall trend is still captured.
%At the panels with $\beta=2.7$ and $\beta=\beta_0$, there are localized abrupt increases that come from higher order resonance, i.e they are obtained at $C^0_K$ values which correspond to $\omega_0=\frac{1}{2}\beta$. For librating KLCs these abrupt increases were mentioned in \cite{Klein2024} but their correspondence to second order resonance (i.e that they occur at $\omega_0=\frac{1}{2}\beta$) was not mentioned. For rotating KLCs there are no first order resonance, i.e there is no value of $C^0_K$ where $\omega_0=\beta$ but the overall trend is reconstructed. 

\section{Discussion}
In this study, we extended the approach presented in \citepaperI\ and \citepaperII\ and analytically study the effect of a constant rate precessing quadrupole potential on high eccentricity Kozai-Lidov Cycles (KLCs). We demonstrated that for precession rates where a first-order resonance occurs, i.e., where there can be a correspondence between the frequency of the KLC and the precession of the perturbing potential, the long-term dynamics can be described using a simple pendulum model. In this model, the angle of the pendulum represents the phase difference between the KLC frequency and the precession of the perturbing potential. This approach allows for the prediction of the amplitude of change in $j_z$, as it is connected to the velocity of the pendulum. Additionally, we have shown that KLCs with $j_z = 0$ are also described by a simple pendulum model (see Appendix \ref{app:pure_Kozai_jz_0_simple_pendulum}).

The analytic development presented in this Letter assumes a close proximity between the precession rate of the perturbing potential and the frequency of the (unperturbed) KLC, as well as the condition that $C_K$ retains its sign throughout the evolution. Under these assumptions, as illustrated in the top panel of Figure \ref{fig:top_omega0_bottom_numeical_deltajz_vs_Ck0_beta_1.8_alpha_1deg}, the applicability is limited to a specific range of precession rates. As discussed in \cite{klein2023,klein2024}, the problem of a precessing quadrupole potential has been analytically studied with a more advanced solution that provides successful approximate predictions for the long-term evolution over a slightly broader range of precession rates. The significance of the solution presented in this Letter lies in its simplicity, as it utilizes a straightforward pendulum model. This simplicity enhances the analytic understanding of the system, providing a more intuitive and deeper insight into the dynamics compared to the more complex models.

Consider, for example, a hierarchical triple system with an inner binary of two m-dwarf stars of $m_1=m_2=0.5M_\odot$ on a circular orbit with semi major axis $a_{12}\sim10\text{AU}$ which is orbited by a $m_3=1M_\odot$ tertiary star on a keplerian orbit with semi major axis of $a_{\text{outer}}=500\text{AU}$ with inclination of $i\sim 20 ^\circ$. If the tertiary star is orbited by a planet on a $a=5\text{AU}$ orbit, the orbit of the planet is perturbed by a precessing quadrupole potential with $\alpha\sim1^\circ$ and $\beta \sim 2$. In the high eccentricity regime of the planetary motion, this system can be analytically investigated using the simple pendulum model.

% In conclusion, our study provides a comprehensive analytical framework for understanding the long-term evolution of high eccentricity orbits in hierarchical three-body systems, emphasizing the critical role of Brown's Hamiltonian. This approach enhances our ability to predict and analyze the dynamical behavior of various astrophysical systems, such as exoplanets and satellite systems, using a simplified yet accurate model.

% \section*{Acknowledgements}

% We thank ...

%%%%%%%%%%%%%%%%%%%%%%%%%%%%%%%%%%%%%%%%%%%%%%%%%%
\section*{Data Availability}

The codes used in this article will be shared on reasonable request.



\bibliographystyle{mnras}
\bibliography{precessing_quadrupole_is_a_pendulum} % if your bibtex file is called example.bib

%\begin{thebibliography}{99}
%\end{thebibliography}

%%%%%%%%%%%%%%%%%%%%%%%%%%%%%%%%%%%%%%%%%%%%%%%%%%

%%%%%%%%%%%%%%%%% APPENDICES %%%%%%%%%%%%%%%%%%%%%

\appendix

\section{Pure KLC at $j_z=0$ is a simple pendulum}\label{app:pure_Kozai_jz_0_simple_pendulum}

For regular KLCs (i.e, when the perturbing potential is not precessing), with the constant $j_z=0$, and using the coordinate system where $j_x=j_z=e_y=0$, Equations \ref{eq:secular_equations} become
\begin{eqnarray}
    % \begin{equation}
        \frac{de_{x}}{d\tau}=&-\frac{9}{4}j_{y}e_{z}\cr
        \frac{dj_{y}}{d\tau}=&+\frac{15}{4}e_{x}e_{z}\cr
        \frac{de_{z}}{d\tau}=&-\frac{3}{2}j_{y}e_{x}.
    % \end{equation}
\end{eqnarray}
Define the variables $x,y,z$ through 
\begin{eqnarray}
e_{x}=&\sqrt{\frac{2}{3}\frac{4}{15}}x\cr
j_{y}=&\sqrt{\frac{2}{3}\frac{4}{9}}y\cr
e_{z}=&\sqrt{\frac{4}{9}\frac{4}{15}}z
\end{eqnarray}
leading to the following equations of motion
\begin{eqnarray}\label{eq:x_y_z_structure_equations}
    \dot{x}=&-yz\cr
    \dot{y}=&xz\cr
    \dot{z}=&-xy,
\end{eqnarray}
having two constants 
\begin{eqnarray}\label{eq:Cx_Cz}
C_{x}=&x^{2}+y^{2}\cr
C_{z}=&z^{2}+y^{2},
\end{eqnarray}
where the constant $C_K$ (Equation \ref{eq:CK}) 
\begin{equation}
    C_K=\frac{8}{45}\left(C_x-C_z\right)=\frac{8}{45}\left(x^2-z^2\right)
\end{equation}
so 
\begin{eqnarray}\label{eq:Cx_Cz_as_function_of_Ck}
C_{x}=&\frac{9}{8}\left(2C_{K}+3\right)\cr
C_{z}=&\frac{9}{8}3\left(1-C_{K}\right).
\end{eqnarray}
The equations \ref{eq:x_y_z_structure_equations} are equations of a simple pendulum through the following correspondence: Consider a simple pendulum with angle $\theta$, i.e $\ddot{\theta}\propto \sin\theta$. Using the following change of variables:
\begin{eqnarray}
    x=&\frac{\dot{\theta}}{2}\cr
    z=&\frac{1}{\sqrt{L}}\cos\left(\frac{\theta}{2}\right)\cr
    y=&\frac{1}{\sqrt{L}}\sin\left(\frac{\theta}{2}\right)
\end{eqnarray}
with $L$ constant, these variables behave dynamically with the structure of Equations \ref{eq:x_y_z_structure_equations} - so KLCs at $j_z=0$ is a simple pendulum with a velocity $\dot{\theta}\propto e_x$. If the pendulum is rotating, $\dot{\theta}$ does not change its sign and $e_x \neq 0$ and since $\theta$ can have any value between $0$ and $2\pi$, $e_z$ can cross zero, i.e $C_K>0$ so rotation of the pendulum is a rotation of the KLC. Alternatively, if the pendulum is librating, its velocity changes sign so $e_x$ can cross zero, i.e $C_K<0$ and libration of the pendulum is a librating KLC. Using the correspondence to a simple pendulum we obtain the period of the KLC:
\subsection{rotation, $C_{z}<C_{x}$, $x \neq 0$}
Since $z$ can cross zero we have $C_{z}=y_{\text{max}}^{2}$. Assign the following change of parameters:
\begin{eqnarray}
    y=&y_{\text{max}}\sin u\cr
    z=&y_{\text{max}}\cos u
\end{eqnarray}
leading to 
\begin{equation}
    \dot{u}=x=\sqrt{C_{x}}\sqrt{1-\frac{y_{\text{max}}^{2}}{C_x}\sin^{2}u}
\end{equation}
so the period of the pendulum is
\begin{equation}\label{eq:period_rotating_pendulum}
    T=4\intop_{0}^{\frac{\pi}{2}}\frac{du}{\dot{u}}=\frac{4}{\sqrt{C_{x}}}K\left(\frac{C_z}{C_x}\right)
\end{equation}
with $K\left(m\right)$ the complete elliptic function of the first kind.
\subsection{libration, $C_{z}>C_{x}$, $z \neq 0$}
Since $x$ can cross zero we have $C_{x}=y_{\text{max}}^{2}$. Assign the following change of parameters:
\begin{eqnarray}
    y=&y_{\text{max}}\sin u\cr
    x=&y_{\text{max}}\cos u
\end{eqnarray}
leading to 
\begin{equation}
    \dot{u}=z=\sqrt{C_{z}}\sqrt{1-\frac{y_{\text{max}}^{2}}{C_z}\sin^{2}u}
\end{equation}
so the period of the pendulum is
\begin{equation}\label{eq:period_librating_pendulum}
    T=4\intop_{0}^{\frac{\pi}{2}}\frac{du}{\dot{u}}=\frac{4}{\sqrt{C_{z}}}K\left(\frac{C_x}{C_z}\right).
\end{equation}

%%%%%%%%%%%%%%%%%%%%%%%%%%%%%%%%%%%%%%%%%%%%%%%%%%

% Don't change these lines
\bsp	% typesetting comment
\label{lastpage}
\end{document}